\documentclass[10pt,preprint2]{aastex}

\usepackage{natbib}
\bibpunct{(}{)}{;}{a}{}{,}

\def\3c{3C\,273}
\def\Lya{Ly$\alpha$}
\def\Lyal{Ly$\alpha$ $\lambda$1216}
\def\nv{N\,{\sc v}}
\def\nvl{N\,{\sc v} $\lambda$1240}
\def\civ{C\,{\sc iv}}
\def\civl{C\,{\sc iv} $\lambda$1549}
\def\heiil{He\,{\sc ii} $\lambda$1640}
\def\kms{\mbox{km\,s$^{-1}$}}

\received{Date}
\revised{Date}
\accepted{Date}

\ccc{Code}
\cpright{Type}{Year}

\shortauthors{S.Paltani \& M.T\"urler}
\shorttitle{Dynamics of the \Lya\ and \civ\ emitting gas in \3c}

\begin{document}
\title{Dynamics of the \Lya\ and \civ\ emitting gas in \3c}
\author{St\'ephane Paltani}
\affil{Harvard-Smithsonian Center for Astrophysics, 60 Garden St., Cambridge, MA 02138, USA}
\email{spaltani@cfa.harvard.edu}

\and

\author{Marc T\"urler}
\affil{INTEGRAL Science Data Centre, ch.\ d'Ecogia 16, CH-1290 Versoix, Switzerland\and
Geneva Observatory, ch. des Maillettes 51, CH-1290 Sauverny, Switzerland}
\email{Marc.Turler@obs.unige.ch}

\begin{abstract}
  In this paper we study the variability properties of the
  Lyman\,$\alpha$ and \civ\ emission lines in \3c\ using archival IUE
  observations. Our data show for the first time the existence of
  variability on time scales of several years. We study the spatial
  distribution and the velocity field of the emitting gas by
  performing detailed analyses on the line variability using
  correlations, 1D and 2D response functions, and principal component
  analysis.  In both lines we find evidence for two components, one
  which has the dynamic properties of gas in Keplerian motion around a
  black hole with a mass of the order of $10^9$\,M$_\odot$, and one
  which is characterized by high, blue-shifted velocities at large
  lag. There is no indication of the presence of optically thick
  emission medium neither in the \Lya, nor in the \civ\ response
  functions.  The component characterized by blue-shifted velocities,
  which is comparatively much stronger in \civ\ than in \Lya, is more
  or less compatible with being the result of gas falling towards the
  central black hole with free-fall acceleration.  We propose however
  that the line emission at high, blue-shifted velocities is better
  explained in terms of entrainment of gas clouds by the jet.  This
  gas is therefore probably collisionally excited as a result of
  heating due to the intense infrared radiation from the jet, which
  would explain the strength of this component in \civ\ relative to
  \Lya. This phenomenon might be a signature of disk-jet interaction.
\end{abstract}

\section{Introduction}
\label{sec:intro}
The presence of broad emission lines is a very distinctive feature of
quasi-stellar objects (QSOs) and Seyfert 1 galaxies. These lines are
emitted by gas located in the vicinity of the central engine
photoionized by the strong ionizing continuum radiation. The large
width of the emission lines requires velocities well in excess of
$10\,000$\,\kms.  Line profile and relative intensity studies provide
very imporant clues on the gas velocity distribution and on the
physical conditions in the line emitting region
\citep[e.g.,][]{SuleEtal-2000-PheBro}.

A huge amount of information can however be gathered by studying the
variability properties of the emission lines.  Because line emission
is driven by the variable continuum from the central engine, the broad
emission lines vary in response to the continuum variations. As
different parts of the line emitting region respond with different
delays depending on their locations, variability can be used as a tool
to map the entire gas phase-space distribution
\citep{BlanMcke-1982-RevMap}. This provides very important insights on
the global dynamics of the system, and much effort has been devoted to
the development of analysis methods of emission line light curves
\citep[e.g.,][]{Diet-1995-BroEmi}. A very important recent result is
the fact that the large emission line widths are to a large extent due
to Keplerian motions around supermassive objects, and that this effect
provides a way to estimate black hole masses
\citep{PeteWand-1999-KepMot}.

\objectname{3C 273} is a quite peculiar radio-loud QSO that exhibits,
depending on the wavelength domain, the properties of both blazars and
Seyfert 1 galaxies \citep{Cour-1998-BriQua}.  It has been intensively
observed throughout the electromagnetic spectrum for more than 3
decades as part of a large collaboration \citep{TuerEtal-1999-ThiYea},
and in particular throughout the mission of the International
Ultraviolet Explorer (IUE) satellite, which can observe at wavelengths
close to the ionizing continuum, and is particularly well adapted to
the study of important emission lines like \Lya\ (for sufficiently
redshifted objects) and \civ.

In this paper, we extend a previous study of the ultraviolet and
optical continuum emission \citep{PaltEtal-1998-BluBum}, and focus on
the \Lya\ and \civ\ emission line variability based on archival IUE
spectra of \3c.  Previous works on \Lya\ variability in \3c\ provided
conflicting results.  While \citet{ObriEtal-1989-BroEmi} and
\citet{ObriHarr-1991-UltEmi} detected significant line variability,
\citet{UlriEtal-1993-TimVar} found no compelling evidence of
variability. These studies used only a part of the entire IUE data
set; we add here about 5 years of data, which doubles the period of
time with dense sampling. With the addition of these new data, \Lya\ 
variability is now clearly detected, and an analysis of \civ\ is even
possible.  Morevover, the sampling is now sufficient to perform
detailed analyses of the light curves, which allow us to attempt to
constrain the geometry and the velocity field of the emitting gas.

Throughout this paper, the quoted wavelengths are observed
wavelengths, while the Doppler velocities and the time lags are
expressed in the rest frame of \3c, i.e. corrected for a redshift of
$z\!=\!0.158$.

\section{Data}
\label{sec:data}
This paper is based on data from the \3c\ database hosted by the
Geneva Observatory\footnote{Web site:
  \url{http://obswww.unige.ch/3c273/}} and presented in
\citet{TuerEtal-1999-ThiYea}. The data that are used here are the IUE
short wavelength (SWP) spectra processed with the IUE Newly Extracted
Spectra (INES) software \citep{RodrEtal-1999-IueIne}. The sample
consists of sparse observations from 1978 until 1985, followed by ten
years with an observation rate of about once every 2 weeks during two
annual observation periods of 3 months, separated by 2 months. Two SWP
observations were often performed consecutively to improve photometric
accuracy. In total, 256 SWP observations have been performed on 142
different days.

\subsection{Spectra selection}
\label{sec:selection}
Since the inclusion of low quality spectra adds noise or even spurious
features to the results, we apply strict selection criteria.  We first
reject the small aperture spectra and those flagged as dubious in
\citet{TuerEtal-1999-ThiYea}, to which must be added the SWP46739LL
spectrum, because of absence of guiding during the exposure.
Considering only the ranges between 1260 and 1560\,\AA\ for \Lya, and
between 1680 and 1980\,\AA\ for \civ, we discard spectra clearly
contaminated by cosmic rays, those with 10 or more saturated spectral
bins at the top of \Lya, as well as those with a signal-to-noise ratio
(S/N) below 15 on average.\footnote{Although it does not satisfy this
  criterion, SWP19731LL was not discarded, because it is the only
  spectrum between mid-1982 and mid-1984.} The last two selection
criteria are similar to the rejection of over- and under-exposed
spectra applied by \citet{ObriEtal-1989-BroEmi} and
\citet{ObriHarr-1991-UltEmi}. To ensure that differences between \Lya\ 
and \civ\ are not due to a different sampling, we use an identical set
of spectra for both lines.

As in those previous two studies, we correct for the IUE wavelength
scale uncertainty by applying small wavelength shifts ($\le 3$\,\AA)
to the 186 remaining spectra. Overlooking this point can introduce
spurious asymmetric line variability, and can change the line profile
in the rms spectrum (see Sect.~\ref{sec:profiles}).  The wavelength
correction has been calculated by placing the Gaussian fitted peak of
the \Lya\ line at its expected rest wavelength of 1407.7\,\AA.
Finally, after averaging all spectra taken within 12 hours, we end up
with a set of 119 good quality SWP spectra of \3c.

\subsection{Line flux extraction}
\label{sec:extraction}
The \Lyal\ flux (including \nvl) is measured relative to a continuum
defined by a straight line fitted to the points in two 50\,\AA\ 
continuum bands at 1260--1310 and 1510--1560\,\AA.  We integrate the
total line flux for Doppler velocities up to 20\,000\,\kms, i.e.\ from
1314 to 1501\,\AA. The \civl\ flux was measured similarly, with the
continuum bands at 1680--1730 and 1940--1980\,\AA.  To avoid strong
contamination by \heiil, we integrate \civ\ only for velocities up to
10\,000\,\kms, i.e. from 1732 to 1851\,\AA, the observed line center
being at $\sim$\,1791.4\,\AA. Unfortunately, this wavelength is nearly
coincident with an IUE reseau mark. In the three spectra taken before
1980, this mark results in a very clear absorption-like feature, that
we correct by interpolating linearly above the affected wavelength
bins.

The line flux uncertainty is calculated by taking into account both
the flux error in each spectral bin and an estimate of the uncertainty
induced by the fit to the continuum. As the resolution of IUE is about
6\,\AA\ and the spectral bins are $\sim$\,1.7\,\AA\ wide, the spectral
bins are not completely independent. We take this into account by
multiplying the uncertainty resulting from the individual flux errors
by $\sqrt{6/1.7}\simeq 1.9$.  It can be very easily seen that the
uncertainty induced by the fit is expressed as
$(\Delta\lambda/2)\,\sqrt{\sigma_1^2\!+\!\sigma_2^2}$, where
$\Delta\lambda=\lambda_2\!-\!\lambda_1$ is the wavelength range of
line integration and $\sigma_1$, $\sigma_2$ are the continuum fit
uncertainties at $\lambda_1$, $\lambda_2$.  We also extract a
continuum flux averaged over the 1250-1300\,\AA\ range.  The same
provision for non-independent spectral bins is taken to estimate the
continuum flux uncertainty.

The average uncertainties obtained with the above method are
respectively 2\,\% for the continuum, and 3.5\,\% for \Lya\ and \civ.
However, a \emph{structure function analysis}
\citep[e.g.,][]{Palt-1999-ConBll} gives average uncertainties
exceeding the above ones by factors 1.6, 1.2, and 2 respectively. This
means that the INES uncertainties are slightly underestimated (which
we have confirmed using a single-bin, continuum light curve).  We
therefore apply these correction factors to our continuum, \Lya, and
\civ\ light curves. We finally obtain average uncertainties of
3.5\,\%, 4.5\,\%, and 7\,\%, respectively.

\subsection{Average and rms line profiles}
\label{sec:profiles}
\begin{figure}[tbp]
\plotone{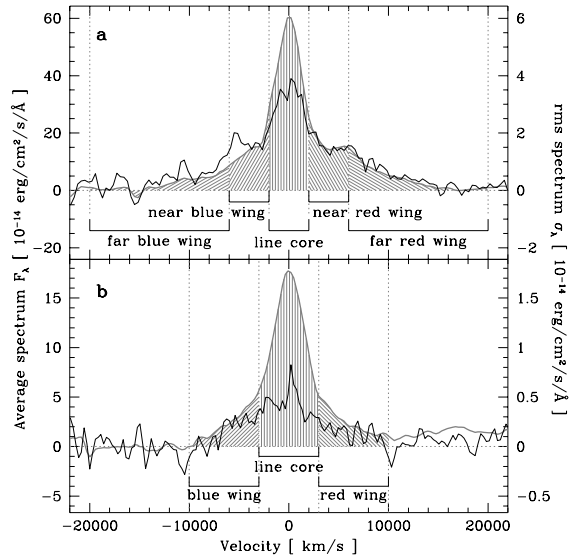}%
\caption{\label{fig:mean_rms}
  Average \Lya\ \textbf{(a)} and \civ\ \textbf{(b)} line (grey solid
  line) and rms profiles (black solid line). For better comparison the
  rms profile is shown on a ten times larger scale. The definition of
  the different parts of the lines are also shown.}
\end{figure}
In Fig.~\ref{fig:mean_rms}a we compare the average \Lya\ emission line
profile and the root-mean-square (rms) profile. The rms profile is
defined as in \citet{WandEtal-1999-CenMas}.  The rms \Lya\ profile is
comparably broader than the average profile, which indicates that the
wings of this line are much more variable than its core. Another
significant difference is an excess in the blue wing variability
(around $-5000$\,\kms). The contamination of \Lya\ by the \nvl\ line
in its red wing (around $+5000$\,\kms) is visible, but relatively
small.  The \civ\ rms profile shows similar features, but with a
lesser S/N.

\subsection{Light curves}
\label{sec:varline}
\begin{figure}[tbp]
\plotone{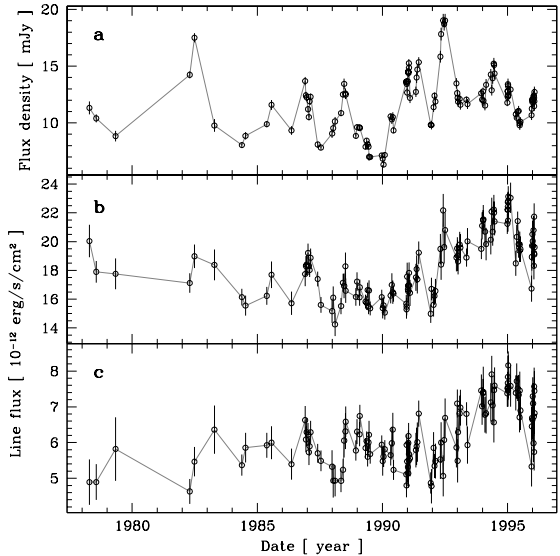}%
\caption{\label{fig:lcurve}
  \textbf{(a)} Ultraviolet continuum light curve averaged in the
  1250--1300\,\AA\ band.  \textbf{(b)} \Lya\ emission line light curve
  integrated for velocities up to 20\,000\,\kms. \textbf{(c)} \civ\ 
  emission line light curve integrated for velocities up to
  10\,000\,\kms.}
\end{figure}

The light curves of the integrated \Lya\ and \civ\ lines and of the
ultraviolet continuum (1250--1300\,\AA) are shown in
Fig.~\ref{fig:lcurve} and printed in Table \ref{tab:lightcurve}
(published in full in the Electronic Version). They can be obtained
from the Geneva Observatory web site
\url{http://obswww.unige.ch/3c273/}.  Both lines appear to be clearly
variable, as confirmed by the large reduced chi-square values of
$\chi_{\nu}^2\,\simeq\,10.3$ for \Lya\ and $\chi_{\nu}^2\,\simeq\,4.6$
for \civ\ obtained under the hypothesis that the line flux is constant
(118 degrees of freedom).

The peak-to-peak variability of the \Lya\ light curve is $\sim$\,50\%
of the mean line flux ($\sim$\,60\% for \civ). This is about three
times the value found by \citet{UlriEtal-1993-TimVar}, whose study
could not benefit from the data obtained after July 1992.  The
relative variability, defined as the ratio of the standard deviation
corrected for the average uncertainty to the mean flux
\citep{ClavEtal-1991-SteTow}, is $0.11$ for \Lya, $0.12$ for \civ, and
$0.21$ for the continuum.

Visual comparison between Figs~\ref{fig:lcurve}a, \ref{fig:lcurve}b,
and \ref{fig:lcurve}c shows that the \Lya\ and \civ\ light curves vary
on significantly longer time scales than the continuum, one huge bump
lasting at least 4 years dominating the line variability. On short
time scales however, the continuum and line light curves are quite
similar, as each peak in the line light curves can be found in the
continuum.

The average post-1992 line flux is significantly higher than that
observed until then. This variability on long time scales could not be
observed in previous studies
\citep[i.e.][]{ObriEtal-1989-BroEmi,ObriHarr-1991-UltEmi,UlriEtal-1993-TimVar},
only based on earlier data. As a consequence, this study can explore
new facets of (in particular) \Lya\ emission properties.

\section{Correlation analysis}
\label{sec:croco}
We perform cross-correlation analyses with our data using two
different methods: the interpolated cross-correlation function method
\citep[ICCF;][]{GaskPete-1987-AccCro} as implemented by
\citet{WhitPete-1994-ComCro}, and the $z$-transformed discrete
correlation function (ZDCF) \citep{Alex-1997-AgnVar}. The presence of
a period with poor sampling (before 1985) can in principle strongly
affect the ICCF because of interpolation. We checked however that the
effect is negligible in our case by comparing the ICCFs of the full
data set with those using only post-1985 data. We therefore keep below
the full set of data for consistency with the other analyses presented
here.

\begin{figure}[tbp]
\plotone{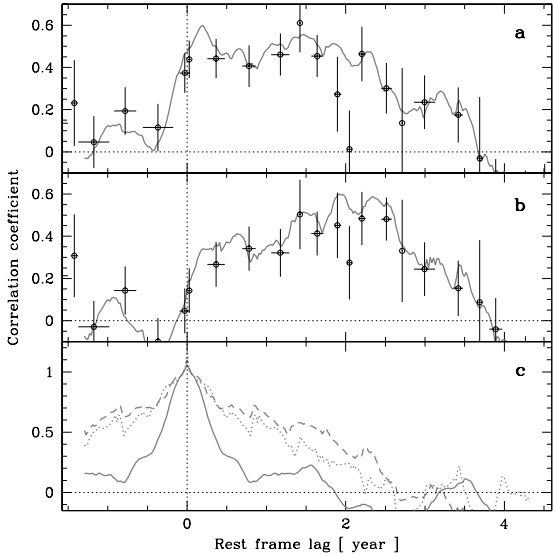}%
\caption{\label{fig:croco}
  \textbf{(a)} and \textbf{(b)} Correlations between the UV
  1250-1300\AA\ continuum emission and the \Lya\ \textbf{(a)}, and
  \civ\ \textbf{(b)} emission lines. The solid line is the ICCF.  The
  points are the ZDCF. \textbf{(c)} ACFs of the continuum (solid
  line), \Lya\ (dashed line), and \civ\ (dotted line) emission lines.
  For better comparisons, ACFs are corrected for the effect of
  uncertainties (see text).}
\end{figure}

Fig.~\ref{fig:croco} shows the correlation between the 1250-1300\,\AA\ 
continuum and the \Lya\ and \civ\ emission lines obtained with the two
methods. The agreement is very good, except for a strange dip at a lag
of slightly more than 2 years in the ZDCF with \Lya. The significance
of the departure of the correlation point at lag $2.36$\,yr from the
ICCF is however only about 2\,$\sigma$.

Qualitatively, the correlations exhibit a rapid increase of the
correlation for $\tau>0$, as expected if the lines are driven by the
UV continuum.  The peaks are however extremely broad, the correlations
returning to zero only at a lag of almost $4$\,yr.
Fig.~\ref{fig:croco} can be compared to Fig.~9c of
\citet{KoraGask-1991-StrKin}. Their cross-correlation drops to zero
after a bit more than one year, much more rapidly than what we
observe. This discrepancy is assuredly the result of the fact that
long-time scale variability was absent (or negligible) in the data set
they have used.  \citet{KaspEtal-2000-RevMea}, whose data set consists
essentially of post-1992 observations, find a closer, but still
smaller, value of about $3$\,yr for H$\alpha$, H$\beta$, and
H$\gamma$. It can be noted however that these low-ionization lines are
expected to lie farther away from the central engine than \Lya\ and
\civ.

An interesting difference between the cross-correlations with \Lya\ 
and \civ\ is that the cross-correlation is markedly lower for lags
smaller than 0.5\,yr in the case of \civ.  In particular, the peak at
0.2\,yr visible in the \Lya\ ICCF is completely absent in the \civ\ 
ICCF. The correlation at 0 lag is about 0.1 for \civ, compared to more
than 0.4 for \Lya. The other parts of the correlations are very
similar in both lines.

If two time series are causally related by a response function (see
Sect.~\ref{sec:response}), the width of the correlation peak is due in
part to the response function, and in part to the autocorrelation of
the driving time series (technically, the correlation is the
convolution of the autocorrelation with the response function). It is
therefore important to test whether the broad correlation peaks in
Figs~\ref{fig:croco}a and b could be due to the autocorrelation of the
ultraviolet continuum. Fig.~\ref{fig:croco}c shows the autocorrelation
functions (ACFs) of the continuum, \Lya, and \civ\ light curves
calculated using the same method as the ICCFs. To get rid of the
effect of uncertainties, which makes comparisons between ACFs
difficult, we normalized the autocorrelations so that they equal $1$
for a very small non-null lag $\tau\!=\!0.04$\,yr. The UV continuum
ACF shows a very narrow central peak with a FWHM of about $0.6$\,yr,
dropping to almost $0$ at $\tau\simeq 0.75$\,yr. This shows that the
correlations are to a large extent dictated by the response function.
Fig.~\ref{fig:croco}c also shows the ACFs of the two lines. They
exhibit much broader FWHMs of the order of $3$\,yr. The \Lya\ ACF is
a little bit broader than that of \civ, but the significance of this
result is questionable.

\begin{figure}[tbp]
\plotone{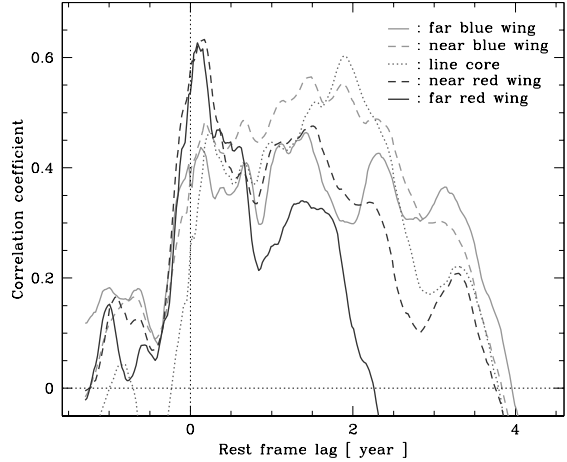}%
\caption{\label{fig:croco5}
  ICCF correlations between the UV 1250-1300\AA\ continuum emission
  and the five parts of the \Lya\ emission line as defined in
  Fig.~\ref{fig:mean_rms}.}
\end{figure}
To gain insight into the dynamics of the line-emitting region, we
decompose the \Lya\ and \civ\ emission lines into respectively 5 and 3
parts as defined in Fig.~\ref{fig:mean_rms}. For \Lya, the line core
is defined as the part of the line with absolute velocities smaller
than $2000$\,\kms. The near wing (either red or blue) is defined by
absolute velocities between $2000$ and $6000$\,\kms. The far wing
(again red or blue) extends from absolute velocities of $6000$\,\kms\ 
up to $20\,000$\,\kms\ (The choice of $6000$\,\kms\ distributes more
or less evenly the contamination by \nv\ in the near and far red
wings).  Because of S/N limitations, we could not use the same
velocity boundaries for \civ.  We define the \civ\ line core as the
part of the line with absolute velocities up to $3000$\,\kms, the rest
being respectively the blue and red wings. Table~\ref{tab:var} shows
the relative variability of the different parts of the \Lya\ and \civ\ 
lines. In both lines, the blue part varies significantly more than the
red one, or that the core.

Fig.~\ref{fig:croco5} shows the ICCFs between the ultraviolet
continuum and the five parts of \Lya\ defined above. A very moderate
smoothing of the curves at large lag has been applied in this figure
for clarity.  The major difference is that both the ZDCF and the ICCF
for the far red wing decrease much faster than those with the other
parts of the line, reaching 0 correlation at a lag close to 2\,yr
instead of 4.  A faster decrease in correlation for the near red wing
is also observed, although it is marginal in the ICCF. These results
indicate that the \Lya\ blue wing varies on time scales significantly
longer than the red wing.  Another difference is that the peak at very
small lag is only present in the red wing.  It must be noted however
that this excess of correlation is not at all compelling in the ZDCFs.

\begin{figure}[tbp]
\plotone{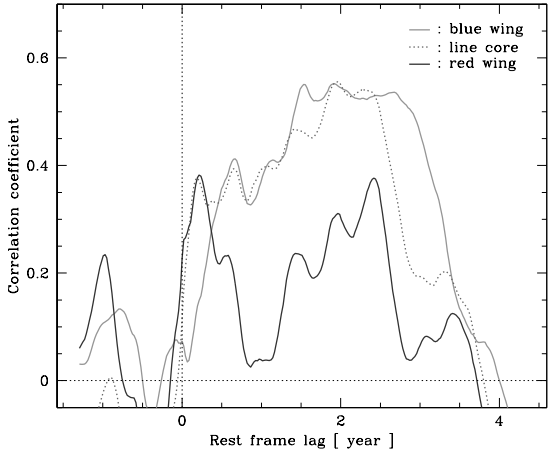}%
\caption{\label{fig:croco3}
  ICCF correlations between the UV 1250-1300\AA\ continuum emission
  and the three parts of the \civ\ emission line as defined in
  Fig.~\ref{fig:mean_rms}.}
\end{figure}
The correlations of the ultraviolet continuum with the different parts
of the \civ\ line (Fig.~\ref{fig:croco3}) also show a faster drop of
correlation in the red wing, although the correlation has a quite
peculiar shape. The much smaller correlation at very small lag is here
also mostly contained in the red wing of the line.

\section{Response function analysis}
\label{sec:response}
\subsection{1D and 2D response functions}
It is well known that the emission line variations in Seyfert galaxies
follow those of the ultraviolet continuum
\citep[e.g.,][]{Pete-1998-RevMap}. This is naturally explained in
models where gas in the periphery of an active galactic nucleus (AGN)
is illuminated by the ultraviolet radiation from the central engine.
As the gas responds to the continuum variation with a delay depending
on its location, we can write the relationship between the overall
emission line flux $F_{\mathrm{L}}(t)$ and the continuum flux
$F_{\mathrm{C}}(t)$ in the form of the following convolution
\citep{KrolDone-1995-RevMap}:
\begin{equation}
  \label{eq:conv}
  F_{\mathrm{L}}(t) = <\!\!F_{\mathrm{L}}\!\!> + \int_{-\infty}^{+\infty}\!\!\Psi(\tau)\cdot
  [F_{\mathrm{C}}(t-\tau)-<\!\!F_{\mathrm{C}}\!\!>]~ \mathrm{d}\tau,
\end{equation}
where $\Psi(\tau)$ is the 1-dimensional response function, depending
on $\tau$ only, and is mostly defined by the geometry of the line
emitting region; and $<\!\!x\!\!>$ denotes the time average of the
variable $x$. Note that the form of Eq.~(\ref{eq:conv}) allows for
arbitrary non-linear relationship between $F_{\mathrm{L}}(t)$ and
$F_{\mathrm{C}}(t-\tau)$, provided it can be locally approximated by a
linear relationship for variations of $F_{\mathrm{C}}$ around
$<\!\!F_{\mathrm{C}}\!\!>$.

Solving Eq.~(\ref{eq:conv}) for $\Psi(\tau)$ can in principle provide
very strong constraints on the BLR geometry.  We perform this
deconvolution using the regularized linear inversion deconvolution
method, first applied to AGN reverberation studies by
\citet{KrolDone-1995-RevMap}, one of the standard methods described by
\citet{Horn-1999-EchMap}: We first transform the deconvolution into a
set of linear equations, which is in general very ill-conditioned.  We
sample the response function with a higher time resolution for
$\tau\le 1$\,yr. To avoid completely unstable solutions, a matrix
imposing the smoothness of the solution is added to the matrix
representing the convolution.  Several kinds of smoothing matrices can
be chosen. We opt here for the simplest one, which imposes a constant
solution on the equation system. By increasing the weight of the
smoothing matrix, one can get arbitrarily smooth solutions. On the
other hand, setting the weight to $0$ removes all smoothing, and the
method becomes equivalent to a singular value decomposition. By
convention, a weight of $1$ means that the trace of the convolution
matrix and that of the smoothing matrix are identical.  Throughout the
paper, we adopt a weight of $5$.  No additional smoothing is applied.
All numerical integrations are made using a simple trapezoidal
formula. Technical details on the method can be found in
\citet{PresEtal-1992-NumRec}.

Eq.~(\ref{eq:conv}) gives the relation between the continuum and the
total line flux.  However, we can alternatively define a response
function for each particular velocity $v$. This gives the
2-dimensional response function $\Psi(\tau, v)$, which relates the
line and continuum fluxes according to the following equation:
\begin{equation}
  \label{eq:conv2d}
  F_{\mathrm{L}}(t,v) = <\!\!F_{\mathrm{L}}(v)\!\!> +\\ \int_{-\infty}^{+\infty}\!\!\Psi(\tau, v)\cdot
  [F_{\mathrm{C}}(t-\tau)-<\!\!F_{\mathrm{C}}\!\!>]~ \mathrm{d}\tau
\end{equation}
2D response functions of BLRs in AGN have been studied theoretically
by several authors
\citep[e.g.,][]{BlanMcke-1982-RevMap,WelsHorn-1991-EchIma,
  PereEtal-1992-Resiii,ObriEtal-1994-ResFun}, but, to our knowledge,
only \citet{WandEtal-1995-GeoKin} (to a limited extent) and
\citet{UlriHorn-1996-MonLif} have attempted to derive them from actual
observations. In practice, we integrate $F_{\mathrm{L}}(t,v)$ over
velocity bins of width 500\,\kms\ close to the center of the line and
up to 2000\,\kms\ in the outer wings, and replace
Eq.~(\ref{eq:conv2d}) by a small number of independent equations
having the form of Eq.~(\ref{eq:conv}).  We use the same regularized
linear inversion method to solve these equations.

\subsection{Results}
\begin{figure}[tbp]
\plotone{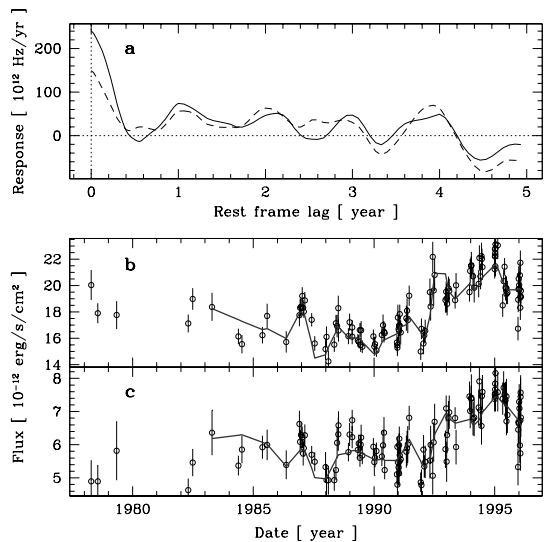}%
\caption{\label{fig:response}
  {\bf (a)} Response function linking the \Lya\ (solid line) and \civ\ 
  (dashed line) emissions to the 1250-1300\,\AA\ UV continuum. The
  \civ\ response has been multiplied by a factor 3. {\bf (b)}
  Comparison between the \Lya\ light curve (points) and the UV
  continuum convolved with the above response function (solid line).
  {\bf (c)} Same as (b) for the \civ\ emission line.}
\end{figure}
Fig.~\ref{fig:response} shows the 1D response functions and the
reconstructed \Lya\ and \civ\ light curves resulting from the
convolution of the UV continuum with the response function. We choose
the smoothing weight so as not to wash out the features in the
response that appear robust (i.e.\ those that are present over a large
range of weights, namely roughly from $1$ to $50$).  The reconstructed
light curve matches reasonably well the original light curves,
although the effect of measurement uncertainties is quite obvious in
the case of \civ. We also checked that the stability of the response
function is very good when its time sampling is modified, and when a
few points are removed from the light curves.

The response functions of both line light curves are mainly
characterized by a quite prominent peak at 0 lag which vanishes for
$\tau\simeq 0.5$\,yr, followed by evidence of significant response up
to $\tau\simeq 4$\,yr. It must be noted that the significance of the
structures in the response function, like the bumps and hollows, is
difficult to assess and will therefore be left out of our discussion.
The \civ\ response function differs mostly from that of \Lya\ in the
ratio between the response at small lags $\tau\le 0.5$\,yr and that at
longer lags $\tau\ge 0.5$\,yr.  This ratio is about twice lower in
\civ\ than in \Lya.

\begin{figure}[tbp]
\plotone{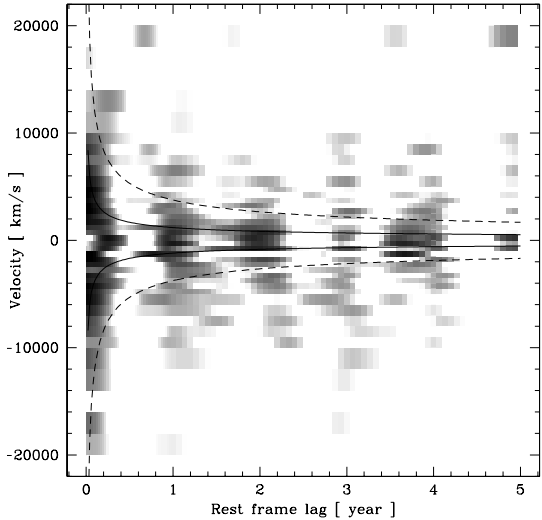}%
\caption{\label{fig:resp2d}
  2D response function of the \Lya\ emission line. Response below a
  given threshold is indicated by a white zone; black zones indicate
  highest response. The lines are the envelopes of the response
  function for a Keplerian circular velocity field around a black hole
  of $10^8$\,M$_\odot$ (solid lines) and $10^9$\,M$_\odot$ (dashed
  lines) respectively.}
\end{figure}
Fig.~\ref{fig:resp2d} shows the 2D response function of the \Lya\ 
emission line (we did not attempt to calculate the 2D response
function for \civ, because of the poorer S/N). The most obvious result
is the very broad response up to $\tau\simeq 0.4$\,yr. The bumps of
Fig.~\ref{fig:response} are clearly visible, and have also markedly
narrower extent in velocity, as expected if their widths are due to
Keplerian motion around a massive object (i.e., $v\sim r^{-1/2}$). A
third feature in Fig.~\ref{fig:resp2d} is a clear excess of response
at lag $\tau\ge 1$\,yr for negative velocities compared to the
positive ones.

\begin{figure}[tbp]
\plotone{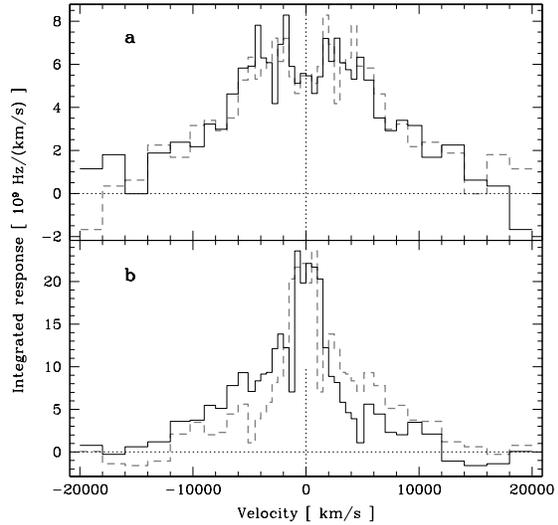}%
\caption{\label{fig:profile}
  \Lya\ line profiles for two lag ranges. {\bf (a)} Profile integrated
  over the range $\tau\le 0.5$\,yr. {\bf (b)} Profile integrated over
  the range $1\le\tau\le 4$\,yr. In order to make the comparison
  between the red and blue sides easier, the grey dashed line in each
  panel represents the mirror of the actual profile (solid line).}
\end{figure}

To assess more quantitatively this latter result, we average
$\Psi(\tau, v)$ over wide ranges of $\tau$, and construct the response
profiles over these ranges. We define a range corresponding to the
central peak ($\tau\le 0.5$\,yr) and one extending from $\tau=1$ up to
$\tau=4$\,yr.  Fig.~\ref{fig:profile} shows the two line profiles
obtained this way.

The short-lag profile in Fig.~\ref{fig:profile}a is very symmetrical.
It is well fitted by a Gaussian with a FWHM of $\sim 18\,000$\,\kms.
If left as a free parameter, the center of the Gaussian is located at
a velocity of $-260$\,\kms, a value not significantly different from
zero.

The long-lag profile in Fig.~\ref{fig:profile}b is however quite
asymmetrical. Two important features emerge from this profile: First
there are large velocities ($\sim 10\,000$\,\kms\ or more)
contributing to the blue wing of the profile. Second, the red wing
drops sharply at velocities of the order of $1500$\,\kms\ already.
The peak around $6000$\,\kms, very close to the location of the \nv\ 
emission line, is probably not significant. We note that a
contamination by \nv\ should rather be observed in the short-lag
profile, as \nv\ should follow the continuum with a smaller lag than
\Lya; its absence in Fig.~\ref{fig:profile}a shows that \nv\ is
largely negligible. Below $1500$\,\kms, the profile is quite
symmetric.

\section{Principal component analysis of the \Lya\ and \civ\ profiles}
\label{sec:pca}
To test further the existence of the two variability modes suggested
by the response function analysis, we performed a {\em principal
  component analysis} \citep[PCA; e.g.][]{KendStua-1976-AdvThe}
independently on the \Lya\ and \civ\ line profiles. Each of the 179
INES bins of the \Lya\ profile between 1260 and 1560\,\AA\ and of the
\civ\ profile between 1680 and 1980\,\AA\ is extracted as a separate
light curve containing 119 measurements. Each of the 119 \Lya\ and
\civ\ profiles can be represented as a point in a 179-dimensional
space (one axis per wavelength bin).  The PCA finds an orthonormal
basis in this space in which most of the coordinates of the 119 points
are small. As a result, by projecting these points on the new basis
vectors, we can represent the observed line profiles by a linear
combination of a few characteristic line profiles determined by the
PCA. In the PCA method, the new orthonormal basis is determined by the
eigenvectors of the covariance matrix of the $119\times 179$ matrix
representing either the \Lya\ or \civ\ profiles at different times.
More details on the method can be found in
\citet{TurlCour-1998-PriCom}.

As the response function indicates the existence of line-emitting gas
responding rather quickly to the continuum, we perform the PCA on the
total spectrum between 1260 and 1560\,\AA\ for \Lya\ and between 1680
and 1980\,\AA\ for \civ\ without subtracting the continuum. In this
way we force the PCA to produce a line profile whose variations match
very well those of the continuum. Any remaining line variability will
therefore, by construction, have a light curve uncorrelated with the
continuum at short lags.
\begin{figure}[tbp]
\plotone{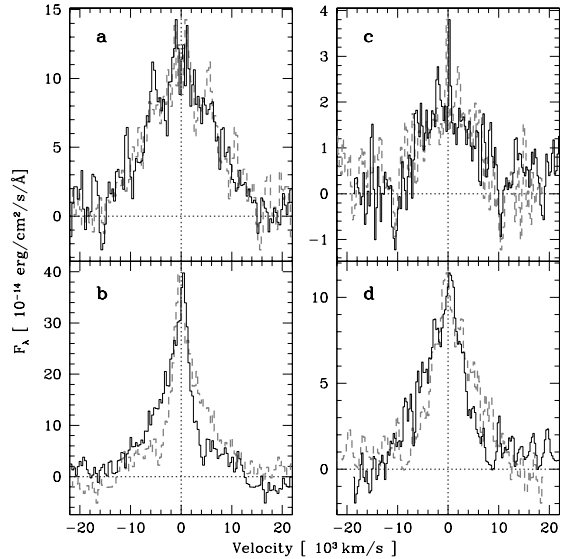}%
\caption{\label{fig:pca}
 \Lya\ and \civ\ variability line profiles as obtained using the PCA. On
  the left, first {\bf (a)} and second {\bf (b)} variability modes for \Lya\ 
  and on the right, first {\bf (c)} and second {\bf (d)} variability modes
  for \civ. The profiles are mirrored as in Fig.~\ref{fig:profile}.}
\end{figure}
Fig.~\ref{fig:pca} shows the profiles (i.e.\ the eigenvectors) of the
two main variability modes.  The profiles of the first eigenvectors of
\Lya\ and \civ\ are both mostly symmetric, broad bumps, with FWHM's of
about $16\,000$ and $14\,000$\,\kms, respectively (the spike in the
\civ\ profile is certainly due to the reseau mark). The profiles of
the two second eigenvectors present on the other hand an asymmetry,
the red wing falling off quite faster than the blue one.  The
similarity of the profiles of Fig~\ref{fig:pca} with those of
Fig.~\ref{fig:profile} obtained using the response function are quite
remarkable, considering that the methods used to derive them are
fundamentally different.  Again, very little contamination by \nv\ is
observed in the \Lya\ profiles, except possibly at about $6000$\,\kms\ 
in Fig.~\ref{fig:pca}b.

\begin{figure}[tbp]
\plotone{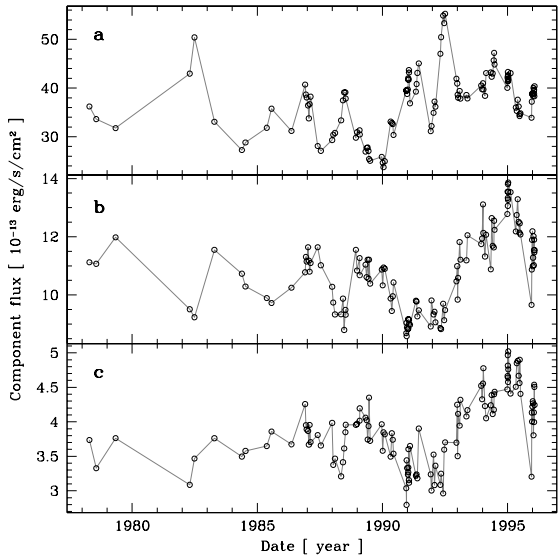}%
\caption{\label{fig:pca_lc}
  Pseudo-light curves associated with the first {\bf (a)} and second
  {\bf (b)} variability modes as obtained with the PCA for \Lya. {\bf
    (c)} Pseudo-light curve associated with the second variability
  mode for \civ.}
\end{figure}

For each variability mode, one can define a pseudo-light curve by
projecting the individual measurements on the axis (in the
179-dimensional space) defined by the eigenvector.
Fig.~\ref{fig:pca_lc} shows three pseudo-light curves obtained this
way for \Lya\ and \civ.
As expected, the first eigenvector for \Lya\ produces a pseudo-light
curve very similar to that of the ultraviolet continuum
(Fig.~\ref{fig:lcurve}a).  The first \civ\ pseudo-light curve is
almost identical to that of \Lya\ and is therefore not shown in
Fig.~\ref{fig:pca_lc}. On the other hand, the pseudo-light curves of
the second eigenvectors for both \Lya\ and \civ\ are characterized by
variations on longer time scales dominated by the large bump between
1992 and 1996.

The remaining 177 eigenvectors found by the PCA have progressively
smaller and smaller associated eigenvalues, which are the variances of
the pseudo-light curves. This means that their associated variability
modes become less and less significant. Most of these minor
variability modes are expected to be white noise due to statistical
fluctuations.  This can be checked using a \emph{structure function}
(SF) \emph{analysis} \citep[e.g.,][]{Palt-1999-ConBll}. We find indeed that
already the third eigenvector pseudo-light curves have a nearly flat
SF, which means that they are most probably dominated by statistical
fluctuations.

\section{Discussion}
\label{sec:disc}
\subsection{General properties}
In this analysis, we have used difficult techniques, some of them (in
particular the 2D response function) pushing the data to their limit.
however, the use of three completely different analysis methods
provide us with a way to cross-validate our results.  In order to
avoid overinterpretation of the data, we discuss below the results
that are found to be consistent across the different analyses. In
general, the very good level of consistency show that we have really
been able to gain information hidden in the light curves.

A first result from the cross-correlation analysis of the full lines,
and from the 1D response function is that line response on very long
time scales up to $4$\,yr is present in both lines.  In the case of
\civ, the response after $\tau\simeq 1$\,yr seems even dominant.  As
noted above, variations on such long time scales have never been
observed in previous studies of line variability in \3c\ 
\citep{KoraGask-1991-StrKin,KaspEtal-2000-RevMea}.  A simplistic,
optically thin, geometrically spherical BLR model with a power-law
emissivity law results in a well-defined response function: The
response function is flat from $\tau=0$ until $\tau=2\cdot
R_{\mathrm{in}}/c$, then decreases more or less rapidly depending on
the relationship assumed between emissivity and distance, and reaches
$0$ for $\tau\ge2\cdot R_{\mathrm{out}}/c$, $R_{\mathrm{in}}$ and
$R_{\mathrm{out}}$ being respectively the inner and outer radius of
the BLR \citep[see, e.g., Fig.~1 in][]{PereEtal-1992-Resiii}.
Disregarding the bumps and hollows in the response functions (which
are of questionable significance), this model requires
$R_{\mathrm{in}}\sim 0.1$\,lyr and $R_{\mathrm{out}} \sim 2$\,lyr. The
\Lya\ response function requires furthermore a BLR covering factor
decreasing with distance with a steep power-law index $\sim 2.5-3$ to
account for the steep decrease after the initial peak. The power-law
index seems however smaller in \civ, as the decrease after the initial
peak is less rapid. Such an index change would require, for instance,
a stratification of metallicity, distant clouds being more metallic
than nearby ones.

In both response functions, the response at $\tau=0$ is the absolute
maximum. If the line emitting medium were optically thick, emission by
clouds located between the ionizing source and the observer would be
strongly suppressed, and there would be very little response between
$\tau=0$ and $\tau \sim 2\cdot R_{\mathrm{in}}/c$. Thus the response
functions indicate that the emitting medium is optically thin.  Other
evidences of the presence of optically thin emission medium have been
pointed out, like variability properties \citep{ShieEtal-1995-OptThi},
or line ratio measurements
\citep{SnedGask-1999-AnaVel,SnedGask-1999-EffAbu}. Further arguments
are presented in \citet{SuleEtal-2000-PheBro}. Note however that
\citet{Math-1982-QuaPan} finds several possible effects making an
optically thick cloud radiate more or less isotropically. It is
unclear however whether these effects are actually realized.

Analyzing the line profiles, we find both evidence for symmetry at
small lags and asymmetry at longer lags.  The PCA analysis shows that
the \Lya\ and \civ\ light curves in Fig.~\ref{fig:lcurve} can be
interpreted as being the result of two distinct variability modes: a
broad and symmetric component of the line responding
quasi-simultaneously to the continuum and an asymmetric component
presenting an excess in the blue wing with slower variations dominated
by a large bump between 1992 and 1996. While the PCA will always
separate a set of $n$ light curves into $n$ distinct components, the
very good match between the profiles obtained with the 2D response
function (Fig.~\ref{fig:profile}) and those obtained with the PCA
(Fig.~\ref{fig:pca}) gives us strong confidence that the total \Lya\ 
and \civ\ lines are really formed by a combination of two separate
components. The same asymmetry at long lags is also evident in the
cross-correlation analysis.

Some asymmetry at small lags in the form of a small excess in the red
wing is also observed in the ICCF cross-correlations. However, their
absence from ZDCF cross-correlations and from the 2D response function
and PCA profiles makes the reality of this effect doubtful.

In the rest of the section, we discuss separately the possible origin
of these two components.

\subsection{Origin of the symmetric component}

The symmetric profile is easily explained in terms of non-organized
velocity field, in particular Keplerian motions around a central black
hole.  The existence of such velocity fields in AGN has been shown by
\citet{PeteWand-1999-KepMot}.  The signature of Keplerian motions in
2D response functions is a symmetric envelope whose width decreases as
$\sqrt{1/\tau}$ (where $\tau$ is the lag).  This general behavior is
apparent in Fig.~\ref{fig:resp2d}, where response at lags $\le
0.5$\,yr occupies a much larger velocity range than that at longer
lags.  Elliptical (i.e., with a continuous distribution of
eccentricities between $0$ and $1$), rather than circular,
trajectories with random inclinations seem to match best our
observations, as there is little response at large velocities for very
small lags in the latter case
\citep{PereEtal-1992-Resiii,WelsHorn-1991-EchIma}.  A Keplerian disk
emission is also often considered as a potential source of emission
lines in AGN.  The accretion disk of \3c\ is very probably close to
face-on.  Superluminal velocities with $\beta_{\mathrm{app}}\simeq 5$
\citep{AbraEtal-1996-VelLif} indeed imply that the angle between the
jet (thought to be perpendicular to the disk) and the line of sight is
at most $10^\circ$. In such situation, the predicted response function
is much flatter than that of a spherical shell \citep[see Fig.~5d
in][]{WelsHorn-1991-EchIma}, which does not match well the observed
response functions.

The theoretical envelopes of a 2D response function for a Keplerian
circular velocity field with random inclinations for two different
black-hole masses ($10^8$ and $10^9$ M$_\odot$) are shown on
Fig.~\ref{fig:resp2d}. These envelopes more or less define the
observable upper and lower limits to the 2D response function of any
cloud ensemble in gravitational equilibrium with random
inclinations.\footnote{While clouds may in general have velocities up
  to the escape velocity, only a small fraction of them can be in this
  situation if the equilibrium condition is to be satisfied.} While
the 2D response function is too noisy to provide an accurate
measurement of the black-hole mass, we note that there is significant
response beyond the $10^8$-M$_\odot$ envelope (in particular at short
lags), which is evidence that a $10^9$-M$_\odot$ black hole might be
required. Because of the disk orientation, disk BLR models would
require a black hole 25 times more massive than circular or elliptic
models, which can probably be ruled out.

\subsection{Origin of the asymmetric component}

The existence of asymmetry requires that the velocity distributions on
both sides of the accretion disk differ.  Taken at face value, the
results of the three different analyses all imply that, under the
assumption that the ultraviolet continuum entirely drives the line
emission, there is a significant excess of line-emitting gas animated
by blue-shifted velocities from 2000 to 10\,000\,\kms\ and responding
with delays of a few years. This corresponds to the prediction for an
infall where cloud velocity increases as clouds approach the black
hole (see Fig.~2 $p\!=\!-0.5$ from \citet{PereEtal-1992-Resiii}; as
the diagram is for outflowing gas, the symmetric around the
$v\!=\!0$ axis must be considered).  Assuming free-fall of
clouds stationary at infinite distance, it is possible to observe a
given velocity for a given black-hole mass at lags 4 times larger than
in the case of a circular velocity field.  Therefore, a 10\,000\,\kms\ 
velocity can be observed at a lag of 2\,yr if the black-hole mass is
about $4\cdot 10^9$ M$_\odot$, which is slightly higher than the
estimate using the symmetric part of the line.  It must be noted
however that infall should produce an excess of response and very
strong variability at short lags for high redshifted velocities, which
is not observed. Indeed, Table~\ref{tab:var} shows that, on the
contrary, the blue wings are much more variable.  If the clouds were
optically thick, this absence of redshifted line emission could be
explained by intra-cloud obscuration masking the emission of the
infalling clouds located between us and the continuum source; however,
for optically thin clouds, this argument is not valid.

The ratio between short- and long-lag response is about twice as large
in \Lya\ compared to \civ. To explain this difference, one would have
to assume that the gas in virial motion has lower \civ\ emissivity
compared to the free-falling gas (due, for instance, to a lower
metallicity). Although this cannot be excluded, we do not see a
natural scenario resulting in such a configuration.

The existence of blue-shifted, high-velocity clouds seems inescapable.
However, the fact that they are found at large lags is only required
under the assumption that all line emission is driven by the
ultraviolet continuum.  \citet{PaltEtal-1998-BluBum} performed a
detailed comparison between the ultraviolet and optical continuum
properties of \3c\ using the same IUE data set as here and a very
intensive optical campaign covering a large fraction of the IUE
campaign. They found that, while presenting strong similarities (all
peaks in the ultraviolet are found in the optical light curve), the
optical light curve presents long-term variations that are absent in
the ultraviolet. They concluded on the basis of spectral and
polarimetric arguments that this slowly varying component (called
$\cal{R}$) may be the synchrotron part of the blazar (synchrotron
optical flares have been detected on several occasions in \3c
\citep{CourEtal-1988-RapInf}). The long-term variability properties of
the line emission and of the $\cal{R}$ component are remarkably
similar \citep[See Fig.~8 in][]{PaltEtal-1998-BluBum}.

If we assume that the slowly-varying line emission and the synchrotron
radiation have the same origin, we readily find an explanation for the
high, blue-shifted velocities: Clouds that pass in the vicinity of the
jet may be entrained by it, creating, because of the geometry of \3c,
an outflow in the direction of the observer (clouds on the other side
of the disk, which should be similarly redshifted, may be masked by
the disk). Similar entrainment has been directly observed by
\citet{VanbEtal-1985-OptEmi} in 3C~277.3.

Line excitation by the jet is probably not due to its ionizing
radiation: The synchrotron component is much weaker than the blue-bump
in the ultraviolet domain, and the Compton component emits
significantly only at much higher energies. The large amount of
infrared radiation emitted by the jet raises however the interesting
possibility that it can be an important source of gas heating, and
therefore of collisional excitation. This idea has been developed by
\citet{MaWill-1998-DoeEve}, and the authors have found strong evidence
that this type of excitation takes place in radio-loud quasars
\citep{MaWill-2001-DisHid}. An important feature of this model is that
\civ\ is much more sensitive to collisional excitation than \Lya,
which explains why the slowly varying component is comparatively much
stronger in \civ\ than in \Lya. The correlation with the ultraviolet
continuum at large lag would either be due to a statistical
coincidence in the blue-bump and synchrotron light curves, or be the
result of an unknown interaction between these two components.

\section{Summary and conclusion}
\label{sec:conc}
We have presented a detailed analysis of the \Lya\ and \civ\ emission
line properties in \3c. Using three different analysis methods, we
find completely consistent results, namely that both \Lya\ and \civ\ 
are correlated with the ultraviolet continuum up to large lags of
$\sim 4$\,yr, that part of the lines follows quite closely the
continuum variations, and that there is a strong excess of line
emission at high, blue-shifted velocities (up to $10\,000$\,\kms) and
large lags (up to 3\,yr at least). The most significant difference
between \Lya\ and \civ\ is that, in the latter, this asymmetric excess
is comparatively much stronger.

The \Lya\ and \civ\ response function are roughly compatible with a
spherically symmetric gas distribution between a minimum distance of
$0.1$\,lyr and a maximum distance of $2$\,lyr. We find evidence of
Keplerian velocity distribution in the 2D response function of \Lya.
It is marginally possible to explain the results using two separate
gas components: one with a virial velocity distribution, and one with
a motion consistent with a free fall. This requires a black-hole mass
of a few $10^9$\,M$_\odot$. In both lines, we find evidence that the
emission medium is optically thin.

We suggest however that a completely different explanation might be
preferred: The high, blue-shifted velocities are the result of
entrainment of the surrounding gas by the jet, and correspond
therefore to outflowing gas.  In this picture, the gas is
collisionally excited as it is heated by the infrared radiation from
the jet. This readily explains the importance of this component in
\civ\ compared to \Lya.

An unanswered question is the relationship between the Seyfert (i.e.,
blue-bump) and blazar (emission from the jet) components in \3c.  The
correlation at large lag between the ultraviolet continuum and the
blue-shifted line component might just be a statistical fluke, or
might be the signature of a disk-jet interaction. \3c\ proves once
again to be a fundamental object for the understanding of the AGN
phenomenon.

\acknowledgements{This work is supported in part (SP) by the Chandra
  X-ray Center under NASA contract NAS8-39073.}

\begin{deluxetable}{lccccccccccc}
  \tabletypesize{\tiny}
  \tablecaption{Light curves of the integrated \Lya\ and \civ\ 
    emission lines. Dates are in decimal year, and fluxes in mJy for
    the continuum and in units of $10^{-12}$ erg\,s$^{-1}$\,cm$^{-2}$
    for the lines.\label{tab:lightcurve}}
  \tablewidth{0pt}
  \tablehead{Date&Continuum&\Lya&\civ&\multicolumn{5}{c}{\Lya}&
    \multicolumn{3}{c}{\civ}\\
    &&Full line&Full line&\multicolumn{2}{c}{Blue
      wing}&Core&\multicolumn{2}{c}{Red wing}&
    Blue wing&Core&Red wing\\
    &&&&Far&Near&&Near&Far}
  \startdata
1978.2888&11.32$\pm$0.57&20.04$\pm$1.13&4.89$\pm$0.64&2.30$\pm$0.52&2.71$\pm$0.33&\phantom{1}8.70$\pm$0.51&3.29$\pm$0.34&3.04$\pm$0.55&0.12$\pm$0.30&4.21$\pm$0.41&0.56$\pm$0.22\\
1978.5627&10.40$\pm$0.36&17.90$\pm$0.76&4.89$\pm$0.50&1.78$\pm$0.33&2.70$\pm$0.24&\phantom{1}8.36$\pm$0.42&2.88$\pm$0.23&2.18$\pm$0.34&0.38$\pm$0.19&3.79$\pm$0.40&0.72$\pm$0.14\\
1979.3376&\phantom{1}8.84$\pm$0.47&17.76$\pm$1.07&5.82$\pm$0.89&1.32$\pm$0.44&2.62$\pm$0.30&\phantom{1}8.78$\pm$0.50&3.05$\pm$0.32&2.00$\pm$0.52&1.27$\pm$0.36&3.92$\pm$0.72&0.62$\pm$0.27\\
1982.3021&14.24$\pm$0.32&17.12$\pm$0.68&4.63$\pm$0.35&1.35$\pm$0.30&2.60$\pm$0.19&\phantom{1}8.11$\pm$0.29&2.86$\pm$0.19&2.20$\pm$0.31&0.54$\pm$0.19&3.65$\pm$0.17&0.44$\pm$0.14\\
1982.4883&17.51$\pm$0.40&18.98$\pm$0.82&5.46$\pm$0.41&0.99$\pm$0.36&2.65$\pm$0.21&\phantom{1}8.62$\pm$0.30&3.13$\pm$0.22&3.59$\pm$0.40&0.43$\pm$0.24&4.27$\pm$0.22&0.76$\pm$0.19\\
1983.2867&\phantom{1}9.77$\pm$0.56&18.38$\pm$1.07&6.36$\pm$0.68&1.38$\pm$0.47&2.59$\pm$0.31&\phantom{1}8.76$\pm$0.47&2.96$\pm$0.33&2.68$\pm$0.54&1.13$\pm$0.36&4.09$\pm$0.43&1.15$\pm$0.30\\
1984.3861&\phantom{1}8.05$\pm$0.26&16.14$\pm$0.58&5.37$\pm$0.30&0.88$\pm$0.24&2.27$\pm$0.15&\phantom{1}7.94$\pm$0.26&2.67$\pm$0.16&2.39$\pm$0.26&1.05$\pm$0.16&3.65$\pm$0.19&0.67$\pm$0.13\\
1984.5287&\phantom{1}8.86$\pm$0.33&15.55$\pm$0.69&5.86$\pm$0.42&0.74$\pm$0.29&2.34$\pm$0.20&\phantom{1}7.71$\pm$0.36&2.71$\pm$0.22&2.05$\pm$0.31&1.33$\pm$0.23&3.82$\pm$0.23&0.71$\pm$0.17\\
1985.3804&\phantom{1}9.89$\pm$0.27&16.23$\pm$0.62&5.92$\pm$0.35&1.34$\pm$0.25&2.42$\pm$0.16&\phantom{1}7.62$\pm$0.27&2.89$\pm$0.18&1.97$\pm$0.27&1.04$\pm$0.18&3.91$\pm$0.18&0.97$\pm$0.14\\
1985.5698&11.59$\pm$0.44&17.70$\pm$0.92&6.00$\pm$0.46&1.32$\pm$0.40&2.42$\pm$0.26&\phantom{1}7.97$\pm$0.43&2.98$\pm$0.27&3.01$\pm$0.43&0.87$\pm$0.27&4.09$\pm$0.24&1.04$\pm$0.21\\
1986.3660&\phantom{1}9.34$\pm$0.38&15.72$\pm$0.81&5.39$\pm$0.44&1.01$\pm$0.35&2.33$\pm$0.23&\phantom{1}7.70$\pm$0.38&2.71$\pm$0.25&1.97$\pm$0.38&0.68$\pm$0.24&3.94$\pm$0.26&0.78$\pm$0.21\\
1986.9123&13.69$\pm$0.36&17.75$\pm$0.84&6.63$\pm$0.39&0.95$\pm$0.35&2.77$\pm$0.21&\phantom{1}8.59$\pm$0.34&3.08$\pm$0.22&2.36$\pm$0.35&1.09$\pm$0.22&4.61$\pm$0.23&0.93$\pm$0.16\\
1986.9422&12.43$\pm$0.31&18.31$\pm$0.69&6.09$\pm$0.36&1.09$\pm$0.29&2.56$\pm$0.22&\phantom{1}9.05$\pm$0.32&3.31$\pm$0.20&2.30$\pm$0.30&0.83$\pm$0.19&4.35$\pm$0.20&0.91$\pm$0.15\\
1986.9751&12.25$\pm$0.32&18.39$\pm$0.69&6.29$\pm$0.40&1.44$\pm$0.30&2.50$\pm$0.19&\phantom{1}8.76$\pm$0.32&3.30$\pm$0.20&2.39$\pm$0.31&0.93$\pm$0.20&4.27$\pm$0.20&1.09$\pm$0.22\\
1987.0322&11.21$\pm$0.29&19.21$\pm$0.66&6.23$\pm$0.32&1.39$\pm$0.27&2.84$\pm$0.18&\phantom{1}8.95$\pm$0.31&3.27$\pm$0.19&2.76$\pm$0.29&0.84$\pm$0.18&4.23$\pm$0.19&1.16$\pm$0.14\\
1987.0619&10.51$\pm$0.30&18.31$\pm$0.62&5.73$\pm$0.36&1.56$\pm$0.27&2.57$\pm$0.18&\phantom{1}8.49$\pm$0.29&3.13$\pm$0.19&2.55$\pm$0.29&0.73$\pm$0.19&4.06$\pm$0.20&0.93$\pm$0.16\\
1987.0913&11.91$\pm$0.28&18.02$\pm$0.47&5.89$\pm$0.26&1.08$\pm$0.20&2.58$\pm$0.13&\phantom{1}8.57$\pm$0.21&3.11$\pm$0.14&2.68$\pm$0.21&0.69$\pm$0.14&4.27$\pm$0.13&0.93$\pm$0.11\\
1987.1331&12.30$\pm$0.31&18.87$\pm$0.65&6.29$\pm$0.32&1.15$\pm$0.27&2.63$\pm$0.18&\phantom{1}8.82$\pm$0.32&3.41$\pm$0.20&2.86$\pm$0.30&0.88$\pm$0.18&4.21$\pm$0.18&1.20$\pm$0.14\\
1987.4168&\phantom{1}8.10$\pm$0.23&17.40$\pm$0.51&5.70$\pm$0.27&1.00$\pm$0.21&2.32$\pm$0.14&\phantom{1}8.53$\pm$0.27&2.84$\pm$0.15&2.71$\pm$0.23&0.73$\pm$0.15&4.06$\pm$0.15&0.91$\pm$0.12\\
1987.5424&\phantom{1}7.83$\pm$0.23&15.60$\pm$0.55&5.49$\pm$0.29&0.76$\pm$0.21&2.29$\pm$0.15&\phantom{1}8.04$\pm$0.27&2.65$\pm$0.16&1.86$\pm$0.24&0.77$\pm$0.16&3.81$\pm$0.16&0.90$\pm$0.12\\
1987.9935&\phantom{1}9.06$\pm$0.39&15.17$\pm$0.81&5.32$\pm$0.46&0.78$\pm$0.34&2.25$\pm$0.23&\phantom{1}7.77$\pm$0.40&2.47$\pm$0.24&1.89$\pm$0.38&0.40$\pm$0.25&4.25$\pm$0.25&0.67$\pm$0.21\\
1988.0369&\phantom{1}9.54$\pm$0.39&16.10$\pm$0.79&4.93$\pm$0.47&0.79$\pm$0.33&2.00$\pm$0.22&\phantom{1}7.75$\pm$0.38&2.84$\pm$0.25&2.72$\pm$0.38&0.57$\pm$0.26&3.66$\pm$0.26&0.70$\pm$0.21\\
1988.1074&10.13$\pm$0.40&14.25$\pm$0.81&4.93$\pm$0.43&0.50$\pm$0.34&1.77$\pm$0.22&\phantom{1}7.43$\pm$0.39&2.61$\pm$0.24&1.95$\pm$0.38&0.48$\pm$0.24&3.73$\pm$0.26&0.72$\pm$0.21\\
1988.3511&10.87$\pm$0.28&15.52$\pm$0.56&4.93$\pm$0.32&0.95$\pm$0.24&2.02$\pm$0.15&\phantom{1}7.51$\pm$0.26&2.61$\pm$0.17&2.44$\pm$0.27&0.49$\pm$0.17&3.61$\pm$0.19&0.83$\pm$0.14\\
1988.4270&12.51$\pm$0.29&17.14$\pm$0.63&5.24$\pm$0.31&1.12$\pm$0.27&2.43$\pm$0.17&\phantom{1}8.00$\pm$0.29&3.00$\pm$0.18&2.59$\pm$0.27&0.66$\pm$0.17&3.80$\pm$0.19&0.77$\pm$0.13\\
1988.4735&13.44$\pm$0.47&16.92$\pm$0.86&6.06$\pm$0.47&1.25$\pm$0.40&2.18$\pm$0.25&\phantom{1}7.61$\pm$0.40&2.97$\pm$0.27&2.91$\pm$0.42&0.67$\pm$0.27&4.13$\pm$0.26&1.26$\pm$0.22\\
1988.5229&12.59$\pm$0.45&18.29$\pm$0.94&6.31$\pm$0.50&1.96$\pm$0.41&2.63$\pm$0.27&\phantom{1}7.89$\pm$0.42&3.21$\pm$0.28&2.61$\pm$0.45&0.80$\pm$0.28&4.25$\pm$0.27&1.26$\pm$0.23\\
1988.5328&12.48$\pm$0.39&16.59$\pm$0.85&6.58$\pm$0.43&0.77$\pm$0.36&2.26$\pm$0.22&\phantom{1}7.73$\pm$0.35&3.04$\pm$0.25&2.78$\pm$0.42&0.98$\pm$0.25&4.38$\pm$0.22&1.22$\pm$0.20\\
1988.9431&\phantom{1}8.85$\pm$0.24&16.16$\pm$0.61&5.78$\pm$0.28&0.92$\pm$0.23&2.21$\pm$0.15&\phantom{1}8.57$\pm$0.28&2.69$\pm$0.17&1.76$\pm$0.33&0.67$\pm$0.16&4.19$\pm$0.16&0.92$\pm$0.12\\
1988.9901&\phantom{1}9.62$\pm$0.28&17.22$\pm$0.65&6.30$\pm$0.35&0.88$\pm$0.27&2.57$\pm$0.17&\phantom{1}8.19$\pm$0.27&2.95$\pm$0.18&2.64$\pm$0.28&0.96$\pm$0.18&4.22$\pm$0.20&1.13$\pm$0.15\\
1989.0926&\phantom{1}9.60$\pm$0.28&16.13$\pm$0.58&6.23$\pm$0.33&0.71$\pm$0.24&2.24$\pm$0.16&\phantom{1}8.49$\pm$0.29&2.78$\pm$0.18&1.91$\pm$0.28&0.86$\pm$0.19&4.27$\pm$0.18&1.09$\pm$0.15\\
1989.0958&\phantom{1}9.54$\pm$0.25&16.82$\pm$0.55&6.74$\pm$0.31&0.77$\pm$0.22&2.28$\pm$0.15&\phantom{1}8.17$\pm$0.28&3.00$\pm$0.17&2.61$\pm$0.25&1.04$\pm$0.20&4.47$\pm$0.16&1.24$\pm$0.13\\
1989.3282&\phantom{1}7.89$\pm$0.23&15.83$\pm$0.48&5.85$\pm$0.28&0.89$\pm$0.20&2.10$\pm$0.14&\phantom{1}8.08$\pm$0.25&2.61$\pm$0.15&2.14$\pm$0.23&0.79$\pm$0.15&4.20$\pm$0.16&0.85$\pm$0.13\\
1989.3694&\phantom{1}8.43$\pm$0.23&15.74$\pm$0.49&6.04$\pm$0.27&0.99$\pm$0.20&2.13$\pm$0.14&\phantom{1}7.93$\pm$0.26&2.70$\pm$0.15&1.98$\pm$0.22&0.76$\pm$0.15&4.25$\pm$0.16&1.03$\pm$0.12\\
1989.4133&\phantom{1}8.11$\pm$0.22&16.63$\pm$0.52&5.60$\pm$0.27&0.96$\pm$0.21&2.39$\pm$0.14&\phantom{1}8.21$\pm$0.26&2.79$\pm$0.15&2.28$\pm$0.22&0.72$\pm$0.14&3.97$\pm$0.16&0.91$\pm$0.12\\
1989.4515&\phantom{1}7.91$\pm$0.23&15.49$\pm$0.54&6.00$\pm$0.28&0.68$\pm$0.22&2.08$\pm$0.14&\phantom{1}7.79$\pm$0.25&2.73$\pm$0.16&2.21$\pm$0.24&0.77$\pm$0.15&4.15$\pm$0.17&1.08$\pm$0.13\\
1989.4696&\phantom{1}7.00$\pm$0.33&16.60$\pm$0.71&6.22$\pm$0.39&1.38$\pm$0.29&2.41$\pm$0.21&\phantom{1}7.94$\pm$0.37&2.57$\pm$0.22&2.31$\pm$0.33&1.03$\pm$0.22&4.29$\pm$0.23&0.89$\pm$0.19\\
1989.5114&\phantom{1}6.99$\pm$0.21&15.36$\pm$0.48&5.68$\pm$0.26&1.20$\pm$0.19&2.11$\pm$0.13&\phantom{1}7.65$\pm$0.25&2.47$\pm$0.14&1.93$\pm$0.22&0.80$\pm$0.14&3.94$\pm$0.15&0.94$\pm$0.12\\
1989.9752&\phantom{1}7.13$\pm$0.22&16.14$\pm$0.52&5.94$\pm$0.26&1.04$\pm$0.20&2.32$\pm$0.14&\phantom{1}7.89$\pm$0.27&2.64$\pm$0.15&2.25$\pm$0.23&0.73$\pm$0.14&4.18$\pm$0.15&1.04$\pm$0.12\\
1990.0139&\phantom{1}6.84$\pm$0.33&15.37$\pm$0.71&5.48$\pm$0.40&1.06$\pm$0.30&2.11$\pm$0.20&\phantom{1}7.52$\pm$0.35&2.54$\pm$0.22&2.12$\pm$0.33&0.55$\pm$0.22&3.85$\pm$0.22&1.08$\pm$0.19\\
1990.0468&\phantom{1}6.32$\pm$0.23&15.56$\pm$0.58&5.61$\pm$0.29&1.02$\pm$0.21&2.17$\pm$0.17&\phantom{1}7.84$\pm$0.35&2.69$\pm$0.18&1.83$\pm$0.24&0.71$\pm$0.15&3.99$\pm$0.18&0.91$\pm$0.11\\
1990.0983&\phantom{1}7.17$\pm$0.22&15.07$\pm$0.49&5.81$\pm$0.26&0.56$\pm$0.19&2.16$\pm$0.14&\phantom{1}7.90$\pm$0.27&2.38$\pm$0.15&2.06$\pm$0.22&1.03$\pm$0.15&3.87$\pm$0.15&0.91$\pm$0.12\\
1990.3301&10.57$\pm$0.26&16.25$\pm$0.54&5.65$\pm$0.30&1.08$\pm$0.24&2.33$\pm$0.15&\phantom{1}7.69$\pm$0.26&2.84$\pm$0.17&2.30$\pm$0.24&0.72$\pm$0.16&3.83$\pm$0.17&1.09$\pm$0.13\\
1990.3771&10.30$\pm$0.41&16.99$\pm$0.82&5.98$\pm$0.46&1.27$\pm$0.36&2.39$\pm$0.24&\phantom{1}7.50$\pm$0.37&3.08$\pm$0.26&2.76$\pm$0.40&0.82$\pm$0.26&4.08$\pm$0.25&1.08$\pm$0.22\\
1990.4148&10.52$\pm$0.41&16.48$\pm$0.81&6.36$\pm$0.47&1.20$\pm$0.35&2.23$\pm$0.24&\phantom{1}7.88$\pm$0.40&3.01$\pm$0.26&2.16$\pm$0.38&1.14$\pm$0.27&4.12$\pm$0.26&1.10$\pm$0.22\\
1990.4448&\phantom{1}9.34$\pm$0.25&16.39$\pm$0.55&5.24$\pm$0.28&1.13$\pm$0.23&2.33$\pm$0.15&\phantom{1}7.96$\pm$0.27&2.87$\pm$0.16&2.10$\pm$0.24&0.60$\pm$0.15&3.80$\pm$0.16&0.83$\pm$0.12\\
1990.9532&13.62$\pm$0.28&15.68$\pm$0.61&5.11$\pm$0.31&1.06$\pm$0.26&2.23$\pm$0.16&\phantom{1}7.50$\pm$0.25&2.92$\pm$0.17&1.97$\pm$0.28&0.66$\pm$0.17&3.49$\pm$0.19&0.96$\pm$0.14\\
1990.9673&13.65$\pm$0.29&15.29$\pm$0.65&4.80$\pm$0.33&0.82$\pm$0.27&2.18$\pm$0.16&\phantom{1}7.79$\pm$0.26&2.77$\pm$0.17&1.72$\pm$0.28&0.66$\pm$0.18&3.28$\pm$0.17&0.85$\pm$0.14\\
1990.9700&13.44$\pm$0.28&15.48$\pm$0.60&5.94$\pm$0.29&1.02$\pm$0.26&2.04$\pm$0.15&\phantom{1}7.48$\pm$0.24&2.77$\pm$0.17&2.17$\pm$0.27&1.10$\pm$0.17&3.85$\pm$0.17&0.98$\pm$0.14\\
1990.9886&12.68$\pm$0.46&17.57$\pm$0.97&5.47$\pm$0.57&1.95$\pm$0.43&2.41$\pm$0.27&\phantom{1}7.60$\pm$0.43&3.09$\pm$0.29&2.52$\pm$0.44&0.76$\pm$0.30&3.81$\pm$0.27&0.90$\pm$0.24\\
1991.0195&13.66$\pm$0.48&16.69$\pm$1.00&5.31$\pm$0.54&1.17$\pm$0.43&2.23$\pm$0.27&\phantom{1}7.81$\pm$0.44&2.61$\pm$0.28&2.86$\pm$0.45&0.67$\pm$0.30&3.71$\pm$0.27&0.94$\pm$0.23\\
1991.0349&14.51$\pm$0.36&16.55$\pm$0.76&5.14$\pm$0.40&1.03$\pm$0.33&2.37$\pm$0.20&\phantom{1}8.01$\pm$0.33&2.94$\pm$0.21&2.21$\pm$0.33&0.52$\pm$0.22&3.69$\pm$0.21&0.93$\pm$0.17\\
1991.0397&14.51$\pm$0.36&16.92$\pm$0.72&5.99$\pm$0.35&1.19$\pm$0.32&2.43$\pm$0.20&\phantom{1}7.82$\pm$0.31&2.92$\pm$0.21&2.56$\pm$0.33&0.99$\pm$0.20&3.83$\pm$0.19&1.16$\pm$0.16\\
1991.0447&14.40$\pm$0.35&16.96$\pm$0.78&6.18$\pm$0.37&1.22$\pm$0.34&2.69$\pm$0.21&\phantom{1}7.81$\pm$0.31&3.07$\pm$0.21&2.16$\pm$0.33&0.80$\pm$0.21&4.14$\pm$0.20&1.24$\pm$0.17\\
1991.0613&15.26$\pm$0.36&16.97$\pm$0.77&5.50$\pm$0.43&0.81$\pm$0.33&2.45$\pm$0.20&\phantom{1}8.05$\pm$0.31&2.93$\pm$0.21&2.74$\pm$0.35&0.55$\pm$0.23&3.82$\pm$0.22&1.14$\pm$0.18\\
1991.0647&14.87$\pm$0.33&17.83$\pm$0.72&5.68$\pm$0.34&1.31$\pm$0.31&2.71$\pm$0.19&\phantom{1}8.02$\pm$0.30&3.06$\pm$0.20&2.73$\pm$0.32&0.89$\pm$0.19&3.75$\pm$0.19&1.04$\pm$0.14\\
1991.1057&12.20$\pm$0.45&16.44$\pm$0.91&5.56$\pm$0.50&1.56$\pm$0.40&2.31$\pm$0.25&\phantom{1}7.56$\pm$0.40&2.73$\pm$0.27&2.28$\pm$0.43&0.72$\pm$0.29&4.07$\pm$0.27&0.76$\pm$0.23\\
1991.3434&12.74$\pm$0.46&17.52$\pm$0.94&5.79$\pm$0.48&1.18$\pm$0.40&2.25$\pm$0.26&\phantom{1}8.16$\pm$0.43&3.24$\pm$0.29&2.69$\pm$0.45&0.59$\pm$0.27&3.93$\pm$0.25&1.26$\pm$0.23\\
1991.3677&14.00$\pm$0.48&18.09$\pm$0.95&5.90$\pm$0.52&1.36$\pm$0.42&2.55$\pm$0.27&\phantom{1}8.39$\pm$0.44&3.08$\pm$0.29&2.71$\pm$0.44&0.82$\pm$0.30&3.87$\pm$0.28&1.21$\pm$0.24\\
1991.4009&14.70$\pm$0.48&17.38$\pm$0.99&6.01$\pm$0.55&1.42$\pm$0.44&2.68$\pm$0.28&\phantom{1}7.95$\pm$0.43&3.17$\pm$0.30&2.15$\pm$0.46&0.94$\pm$0.30&3.79$\pm$0.28&1.28$\pm$0.24\\
1991.4581&15.35$\pm$0.44&19.23$\pm$0.77&6.81$\pm$0.37&1.88$\pm$0.35&2.76$\pm$0.21&\phantom{1}8.28$\pm$0.33&3.43$\pm$0.22&2.88$\pm$0.34&1.14$\pm$0.21&4.37$\pm$0.19&1.30$\pm$0.17\\
1991.9333&\phantom{1}9.83$\pm$0.26&14.99$\pm$0.64&4.87$\pm$0.29&0.85$\pm$0.26&2.10$\pm$0.17&\phantom{1}7.11$\pm$0.29&2.53$\pm$0.17&2.40$\pm$0.27&0.68$\pm$0.17&3.52$\pm$0.16&0.67$\pm$0.13\\
1991.9656&\phantom{1}9.80$\pm$0.39&16.71$\pm$0.78&4.78$\pm$0.48&1.58$\pm$0.34&2.60$\pm$0.23&\phantom{1}7.56$\pm$0.37&2.86$\pm$0.25&2.11$\pm$0.38&0.71$\pm$0.27&3.36$\pm$0.25&0.71$\pm$0.22\\
1992.0479&11.38$\pm$0.42&15.61$\pm$0.82&5.85$\pm$0.44&0.63$\pm$0.35&2.42$\pm$0.25&\phantom{1}7.45$\pm$0.39&2.78$\pm$0.26&2.32$\pm$0.40&0.85$\pm$0.27&3.95$\pm$0.24&1.05$\pm$0.21\\
1992.0857&12.39$\pm$0.30&16.22$\pm$0.61&5.35$\pm$0.34&0.97$\pm$0.26&2.26$\pm$0.17&\phantom{1}7.80$\pm$0.29&2.72$\pm$0.18&2.46$\pm$0.29&0.94$\pm$0.19&3.58$\pm$0.19&0.82$\pm$0.15\\
1992.1243&11.88$\pm$0.43&16.57$\pm$0.84&5.48$\pm$0.49&1.34$\pm$0.37&2.24$\pm$0.25&\phantom{1}7.59$\pm$0.40&2.92$\pm$0.26&2.49$\pm$0.41&0.89$\pm$0.28&3.85$\pm$0.26&0.74$\pm$0.22\\
1992.3179&15.84$\pm$0.52&19.49$\pm$1.05&5.53$\pm$0.55&2.43$\pm$0.48&2.75$\pm$0.31&\phantom{1}8.08$\pm$0.47&3.41$\pm$0.32&2.82$\pm$0.48&0.98$\pm$0.31&3.75$\pm$0.28&0.80$\pm$0.24\\
1992.3456&17.83$\pm$0.56&18.41$\pm$1.07&6.00$\pm$0.57&1.09$\pm$0.49&2.71$\pm$0.32&\phantom{1}8.25$\pm$0.47&3.28$\pm$0.32&3.07$\pm$0.51&0.80$\pm$0.33&4.01$\pm$0.30&1.18$\pm$0.25\\
1992.4297&19.03$\pm$0.59&22.17$\pm$1.15&5.07$\pm$0.57&2.03$\pm$0.53&3.22$\pm$0.34&\phantom{1}9.01$\pm$0.52&4.10$\pm$0.36&3.80$\pm$0.54&0.74$\pm$0.33&3.67$\pm$0.30&0.66$\pm$0.25\\
1992.4713&18.71$\pm$0.57&19.62$\pm$1.15&6.07$\pm$0.59&1.93$\pm$0.53&2.84$\pm$0.33&\phantom{1}8.53$\pm$0.49&3.53$\pm$0.34&2.79$\pm$0.52&0.83$\pm$0.33&4.33$\pm$0.30&0.91$\pm$0.25\\
1992.5115&19.02$\pm$0.59&20.80$\pm$1.21&6.69$\pm$0.54&1.77$\pm$0.54&3.03$\pm$0.35&\phantom{1}8.94$\pm$0.52&3.75$\pm$0.35&3.30$\pm$0.60&0.71$\pm$0.32&4.50$\pm$0.31&1.48$\pm$0.25\\
1992.9623&13.48$\pm$0.46&18.89$\pm$0.95&5.86$\pm$0.56&1.58$\pm$0.41&2.66$\pm$0.27&\phantom{1}8.61$\pm$0.44&3.27$\pm$0.29&2.77$\pm$0.45&0.91$\pm$0.31&4.08$\pm$0.29&0.87$\pm$0.24\\
1992.9939&12.62$\pm$0.33&19.53$\pm$0.77&7.09$\pm$0.39&1.63$\pm$0.32&3.21$\pm$0.21&\phantom{1}8.58$\pm$0.32&3.21$\pm$0.21&2.90$\pm$0.34&1.21$\pm$0.22&4.71$\pm$0.22&1.18$\pm$0.18\\
1993.0104&12.21$\pm$0.50&19.02$\pm$1.06&5.49$\pm$0.61&1.75$\pm$0.45&2.87$\pm$0.30&\phantom{1}8.06$\pm$0.46&3.07$\pm$0.31&3.27$\pm$0.52&0.76$\pm$0.34&4.04$\pm$0.31&0.69$\pm$0.27\\
1993.0260&11.87$\pm$0.43&18.61$\pm$0.89&6.29$\pm$0.54&1.30$\pm$0.38&2.97$\pm$0.27&\phantom{1}8.29$\pm$0.42&3.03$\pm$0.27&3.03$\pm$0.42&1.03$\pm$0.28&4.33$\pm$0.37&0.94$\pm$0.21\\
1993.0876&12.12$\pm$0.43&19.79$\pm$0.88&6.81$\pm$0.50&1.61$\pm$0.39&2.96$\pm$0.27&\phantom{1}9.12$\pm$0.43&3.49$\pm$0.28&2.61$\pm$0.42&1.32$\pm$0.28&4.28$\pm$0.30&1.22$\pm$0.23\\
1993.1206&11.59$\pm$0.43&19.58$\pm$0.94&6.98$\pm$0.50&1.60$\pm$0.39&3.03$\pm$0.27&\phantom{1}8.77$\pm$0.45&3.02$\pm$0.28&3.15$\pm$0.45&1.38$\pm$0.29&4.49$\pm$0.26&1.11$\pm$0.23\\
1993.3609&12.05$\pm$0.28&18.89$\pm$0.64&6.80$\pm$0.33&1.32$\pm$0.27&2.91$\pm$0.18&\phantom{1}8.76$\pm$0.28&3.32$\pm$0.18&2.58$\pm$0.28&1.07$\pm$0.18&4.45$\pm$0.18&1.29$\pm$0.15\\
1993.4022&11.66$\pm$0.43&20.02$\pm$0.92&5.93$\pm$0.52&1.48$\pm$0.39&3.04$\pm$0.27&\phantom{1}9.17$\pm$0.44&3.46$\pm$0.28&2.87$\pm$0.43&0.85$\pm$0.30&4.35$\pm$0.28&0.73$\pm$0.23\\
1993.9539&12.61$\pm$0.36&19.50$\pm$0.71&7.45$\pm$0.45&1.51$\pm$0.31&3.20$\pm$0.22&\phantom{1}9.07$\pm$0.33&3.41$\pm$0.22&2.32$\pm$0.35&1.37$\pm$0.25&4.74$\pm$0.25&1.34$\pm$0.20\\
1993.9889&12.05$\pm$0.44&21.10$\pm$0.94&7.02$\pm$0.52&1.72$\pm$0.40&3.38$\pm$0.28&\phantom{1}9.12$\pm$0.42&3.57$\pm$0.28&3.30$\pm$0.44&1.23$\pm$0.29&4.60$\pm$0.30&1.19$\pm$0.24\\
1994.0226&12.78$\pm$0.33&21.48$\pm$0.71&7.38$\pm$0.40&1.81$\pm$0.30&3.57$\pm$0.21&\phantom{1}9.79$\pm$0.35&3.70$\pm$0.22&2.61$\pm$0.32&1.30$\pm$0.22&4.95$\pm$0.20&1.13$\pm$0.17\\
1994.0377&11.96$\pm$0.53&21.54$\pm$1.01&7.42$\pm$0.71&2.62$\pm$0.47&3.29$\pm$0.31&\phantom{1}9.22$\pm$0.45&3.42$\pm$0.32&3.00$\pm$0.51&1.06$\pm$0.35&4.92$\pm$0.34&1.44$\pm$0.47\\
1994.1039&11.54$\pm$0.44&20.69$\pm$0.98&6.78$\pm$0.50&2.27$\pm$0.41&3.19$\pm$0.28&\phantom{1}8.85$\pm$0.44&3.29$\pm$0.28&3.10$\pm$0.45&1.16$\pm$0.28&4.53$\pm$0.28&1.09$\pm$0.24\\
1994.1477&13.36$\pm$0.50&19.83$\pm$1.13&6.82$\pm$0.56&1.94$\pm$0.48&3.32$\pm$0.33&\phantom{1}9.45$\pm$0.51&3.40$\pm$0.33&1.71$\pm$0.50&1.53$\pm$0.33&4.40$\pm$0.31&0.89$\pm$0.25\\
1994.3354&14.26$\pm$0.49&20.13$\pm$1.03&7.12$\pm$0.54&1.25$\pm$0.44&3.00$\pm$0.29&\phantom{1}8.89$\pm$0.45&3.68$\pm$0.31&3.32$\pm$0.48&1.25$\pm$0.30&4.69$\pm$0.31&1.18$\pm$0.24\\
1994.3687&12.88$\pm$0.46&22.09$\pm$0.97&7.91$\pm$0.52&2.41$\pm$0.42&3.37$\pm$0.29&\phantom{1}9.69$\pm$0.47&3.58$\pm$0.30&3.03$\pm$0.45&1.72$\pm$0.30&4.71$\pm$0.27&1.48$\pm$0.24\\
1994.4094&13.94$\pm$0.49&20.69$\pm$0.98&7.04$\pm$0.53&2.15$\pm$0.44&3.34$\pm$0.30&\phantom{1}9.28$\pm$0.48&3.25$\pm$0.30&2.68$\pm$0.45&1.21$\pm$0.31&4.56$\pm$0.29&1.27$\pm$0.24\\
1994.4534&15.23$\pm$0.51&22.23$\pm$1.03&6.57$\pm$0.56&2.65$\pm$0.47&3.58$\pm$0.31&\phantom{1}9.34$\pm$0.46&3.56$\pm$0.31&3.10$\pm$0.47&1.07$\pm$0.32&4.55$\pm$0.30&0.95$\pm$0.25\\
1994.4676&15.10$\pm$0.37&22.01$\pm$0.79&7.47$\pm$0.38&1.94$\pm$0.34&3.45$\pm$0.23&\phantom{1}9.92$\pm$0.36&3.71$\pm$0.23&2.99$\pm$0.36&1.47$\pm$0.22&4.75$\pm$0.20&1.25$\pm$0.17\\
1994.4885&14.35$\pm$0.49&21.40$\pm$0.97&7.59$\pm$0.51&1.84$\pm$0.43&3.61$\pm$0.31&\phantom{1}9.60$\pm$0.48&3.31$\pm$0.30&3.03$\pm$0.45&1.33$\pm$0.31&4.78$\pm$0.28&1.48$\pm$0.24\\
1994.9970&11.78$\pm$0.30&21.24$\pm$0.71&7.38$\pm$0.37&1.92$\pm$0.29&3.34$\pm$0.20&\phantom{1}9.54$\pm$0.33&3.69$\pm$0.21&2.75$\pm$0.31&1.52$\pm$0.21&4.70$\pm$0.20&1.16$\pm$0.16\\
1995.0084&12.81$\pm$0.21&22.21$\pm$0.54&7.53$\pm$0.26&2.14$\pm$0.21&3.52$\pm$0.14&10.22$\pm$0.23&3.57$\pm$0.14&2.76$\pm$0.23&1.44$\pm$0.14&4.95$\pm$0.14&1.15$\pm$0.11\\
1995.0122&12.31$\pm$0.26&22.75$\pm$0.58&7.50$\pm$0.32&2.55$\pm$0.24&3.61$\pm$0.17&10.04$\pm$0.27&3.56$\pm$0.17&2.98$\pm$0.27&1.31$\pm$0.18&5.09$\pm$0.17&1.10$\pm$0.15\\
1995.0170&12.76$\pm$0.27&22.24$\pm$0.58&7.67$\pm$0.31&2.06$\pm$0.25&3.63$\pm$0.18&\phantom{1}9.86$\pm$0.28&3.69$\pm$0.18&3.01$\pm$0.27&1.46$\pm$0.18&5.06$\pm$0.16&1.15$\pm$0.16\\
1995.0187&13.41$\pm$0.51&22.43$\pm$1.03&7.84$\pm$0.60&2.26$\pm$0.45&3.46$\pm$0.32&10.46$\pm$0.53&3.67$\pm$0.33&2.58$\pm$0.48&1.55$\pm$0.34&4.99$\pm$0.32&1.29$\pm$0.26\\
1995.0243&13.25$\pm$0.35&21.46$\pm$0.76&7.46$\pm$0.45&1.92$\pm$0.33&3.41$\pm$0.22&10.02$\pm$0.37&3.49$\pm$0.22&2.62$\pm$0.34&1.36$\pm$0.23&4.84$\pm$0.22&1.26$\pm$0.29\\
1995.0311&12.45$\pm$0.26&22.82$\pm$0.54&7.55$\pm$0.30&2.33$\pm$0.24&3.75$\pm$0.16&10.03$\pm$0.25&3.62$\pm$0.16&3.09$\pm$0.25&1.29$\pm$0.17&5.00$\pm$0.16&1.26$\pm$0.14\\
1995.0332&12.41$\pm$0.33&23.13$\pm$0.72&8.16$\pm$0.39&2.43$\pm$0.31&3.69$\pm$0.22&10.32$\pm$0.36&3.97$\pm$0.22&2.73$\pm$0.31&1.45$\pm$0.22&5.20$\pm$0.21&1.52$\pm$0.18\\
1995.1223&12.94$\pm$0.49&23.05$\pm$1.05&7.58$\pm$0.58&2.65$\pm$0.47&3.65$\pm$0.32&\phantom{1}9.98$\pm$0.50&3.64$\pm$0.32&3.14$\pm$0.48&0.99$\pm$0.33&4.81$\pm$0.30&1.78$\pm$0.27\\
1995.3352&10.76$\pm$0.43&18.49$\pm$0.86&7.39$\pm$0.49&1.90$\pm$0.38&3.10$\pm$0.27&\phantom{1}9.01$\pm$0.43&3.13$\pm$0.27&1.35$\pm$0.40&1.57$\pm$0.29&4.63$\pm$0.29&1.19$\pm$0.23\\
1995.3735&10.45$\pm$0.42&20.33$\pm$0.85&7.72$\pm$0.49&1.88$\pm$0.37&3.05$\pm$0.27&\phantom{1}9.35$\pm$0.44&3.52$\pm$0.27&2.53$\pm$0.39&1.33$\pm$0.30&4.96$\pm$0.28&1.43$\pm$0.23\\
1995.4120&10.99$\pm$0.30&21.43$\pm$0.66&7.52$\pm$0.35&2.09$\pm$0.27&3.43$\pm$0.20&\phantom{1}9.75$\pm$0.33&3.42$\pm$0.20&2.74$\pm$0.30&1.41$\pm$0.20&4.97$\pm$0.20&1.14$\pm$0.16\\
1995.4498&11.05$\pm$0.44&19.83$\pm$0.95&7.33$\pm$0.56&1.72$\pm$0.40&3.53$\pm$0.29&\phantom{1}9.14$\pm$0.45&3.07$\pm$0.27&2.37$\pm$0.42&1.67$\pm$0.31&4.73$\pm$0.28&0.93$\pm$0.24\\
1995.4854&\phantom{1}9.98$\pm$0.36&19.36$\pm$0.77&6.71$\pm$0.42&1.31$\pm$0.32&3.32$\pm$0.24&\phantom{1}8.79$\pm$0.39&3.38$\pm$0.25&2.57$\pm$0.36&1.00$\pm$0.24&4.73$\pm$0.23&0.97$\pm$0.19\\
1995.4879&\phantom{1}9.80$\pm$0.42&19.80$\pm$0.89&7.46$\pm$0.53&2.01$\pm$0.38&3.16$\pm$0.26&\phantom{1}9.16$\pm$0.41&3.03$\pm$0.26&2.45$\pm$0.42&1.33$\pm$0.30&5.05$\pm$0.28&1.07$\pm$0.24\\
1995.5232&10.09$\pm$0.41&19.47$\pm$0.88&6.90$\pm$0.55&1.93$\pm$0.37&3.17$\pm$0.27&\phantom{1}8.90$\pm$0.43&3.19$\pm$0.27&2.28$\pm$0.41&1.37$\pm$0.35&4.49$\pm$0.29&1.04$\pm$0.23\\
1995.9640&10.68$\pm$0.43&16.73$\pm$0.90&5.33$\pm$0.56&1.51$\pm$0.39&2.64$\pm$0.26&\phantom{1}7.45$\pm$0.39&3.00$\pm$0.27&2.13$\pm$0.41&1.16$\pm$0.29&3.52$\pm$0.31&0.65$\pm$0.23\\
1995.9718&11.25$\pm$0.44&18.93$\pm$0.96&6.45$\pm$0.55&2.05$\pm$0.40&3.02$\pm$0.28&\phantom{1}8.36$\pm$0.37&3.07$\pm$0.27&2.42$\pm$0.45&1.04$\pm$0.28&4.41$\pm$0.36&1.00$\pm$0.24\\
1995.9851&12.05$\pm$0.48&19.71$\pm$1.05&6.69$\pm$0.61&1.22$\pm$0.45&3.07$\pm$0.29&\phantom{1}9.12$\pm$0.46&3.06$\pm$0.30&3.24$\pm$0.49&1.55$\pm$0.33&4.30$\pm$0.37&0.84$\pm$0.25\\
1995.9984&11.82$\pm$0.52&20.51$\pm$0.96&6.64$\pm$0.53&2.07$\pm$0.41&3.13$\pm$0.29&\phantom{1}9.26$\pm$0.45&3.57$\pm$0.29&2.47$\pm$0.43&1.37$\pm$0.30&4.43$\pm$0.28&0.85$\pm$0.24\\
1996.0118&12.05$\pm$0.46&20.04$\pm$0.99&7.29$\pm$0.53&1.86$\pm$0.42&3.14$\pm$0.29&\phantom{1}8.63$\pm$0.43&3.54$\pm$0.30&2.87$\pm$0.45&1.39$\pm$0.30&4.55$\pm$0.30&1.35$\pm$0.24\\
1996.0255&11.93$\pm$0.45&19.71$\pm$0.91&7.10$\pm$0.56&2.15$\pm$0.40&3.01$\pm$0.28&\phantom{1}8.72$\pm$0.42&3.17$\pm$0.28&2.66$\pm$0.43&1.43$\pm$0.30&4.48$\pm$0.31&1.19$\pm$0.24\\
1996.0420&12.33$\pm$0.46&20.79$\pm$0.97&5.98$\pm$0.53&2.49$\pm$0.43&3.09$\pm$0.28&\phantom{1}9.23$\pm$0.46&3.45$\pm$0.30&2.53$\pm$0.44&1.24$\pm$0.30&4.17$\pm$0.30&0.56$\pm$0.24\\
1996.0585&12.42$\pm$0.46&18.32$\pm$0.94&6.73$\pm$0.55&1.79$\pm$0.41&2.94$\pm$0.28&\phantom{1}8.53$\pm$0.42&3.28$\pm$0.29&1.77$\pm$0.44&1.40$\pm$0.30&4.33$\pm$0.30&0.99$\pm$0.24\\
1996.0638&11.48$\pm$0.44&21.73$\pm$0.93&5.74$\pm$0.52&3.17$\pm$0.43&3.58$\pm$0.29&\phantom{1}8.58$\pm$0.44&3.65$\pm$0.30&2.75$\pm$0.42&1.00$\pm$0.29&4.21$\pm$0.29&0.53$\pm$0.23\\
1996.0693&12.20$\pm$0.46&19.20$\pm$0.96&7.57$\pm$0.53&1.59$\pm$0.42&3.16$\pm$0.29&\phantom{1}8.78$\pm$0.45&3.60$\pm$0.30&2.07$\pm$0.44&1.42$\pm$0.30&4.75$\pm$0.29&1.40$\pm$0.24\\
1996.0748&11.91$\pm$0.45&19.11$\pm$0.96&7.45$\pm$0.55&1.60$\pm$0.42&3.01$\pm$0.28&\phantom{1}8.80$\pm$0.44&3.20$\pm$0.29&2.50$\pm$0.43&1.71$\pm$0.30&4.63$\pm$0.31&1.11$\pm$0.24\\
1996.0807&12.72$\pm$0.47&19.61$\pm$0.99&6.82$\pm$0.58&2.04$\pm$0.43&3.13$\pm$0.29&\phantom{1}8.99$\pm$0.46&3.33$\pm$0.30&2.11$\pm$0.46&1.26$\pm$0.31&4.56$\pm$0.33&1.00$\pm$0.25\\
  \enddata
\end{deluxetable}

\begin{deluxetable}{lccccc}
  \tablecaption{Relative variability of the different parts of \Lya\ and \civ.\label{tab:var}}
  \tablewidth{0pt}
  \tablehead{Line&\multicolumn{2}{c}{Blue wing}&Core&\multicolumn{2}{c}{Red wing}\\
  &Far&Near&&Near&Far}
  \startdata
  \Lya&0.288&0.156&0.079&0.086&0.093\\
  \civ&\multicolumn{2}{c}{0.224}&0.083&\multicolumn{2}{c}{0.134}\\
  \enddata
\end{deluxetable}

\end{document}